\documentclass[11pt]{article}

\usepackage[margin=1in]{geometry}
\usepackage{times}        
\usepackage{graphicx}
\usepackage{amsmath,amssymb}
\usepackage{booktabs}
\usepackage{hyperref}

\AtBeginDocument{%
  }

\usepackage{booktabs}

\author{
\begin{tabular}[t]{c@{\hspace{3em}}c}
Timothy Bickmore & Mehdi Arjmand \\
Northeastern University & Northeastern University \\
\texttt{t.bickmore@neu.edu} & \texttt{arjmand.me@northeastern.edu}
\end{tabular}
\\[3em]
Yunus Terzioglu \\
Northeastern University \\
\texttt{terzioglu.y@northeastern.edu}
}

\date{}

\begin{document}

\title{Relational Appliances: A Robot in the Refrigerator for Home-Based Health Promotion}




\maketitle


\begin{abstract}
  Kitchen appliances are frequently used domestic artifacts situated at the point of everyday dietary decision making, making them a promising but underexplored site for health promotion. We explore the concept of relational appliances: everyday household devices designed as embodied social actors that engage users through ongoing, personalized interaction. We focus on the refrigerator, whose unique affordances, including a fixed, sensor-rich environment, private interaction space, and close coupling to food items, support contextualized, conversational engagement during snack choices. We present an initial exploration of this concept through a pilot study deploying an anthropomorphic robotic head inside a household refrigerator. In a home-lab apartment, participants repeatedly retrieved snacks during simulated TV ``commercial breaks'' while interacting with a human-sized robotic head. Participants were randomized to either a health-promotion condition, in which the robot made healthy snack recommendations, or a social-chat control condition. Outcomes included compliance with recommendations, nutritional quality of selected snacks, and psychosocial measures related to acceptance of the robot. Results suggest that participants found the robot persuasive, socially engaging, and increasingly natural over time, often describing it as helpful, aware, and companionable. Most participants reported greater awareness of their snack decisions and expressed interest in having such a robot in their own home. We discuss implications for designing relational appliances that leverage anthropomorphism, trust, and long-term human–technology relationships for home-based health promotion.
\end{abstract}

\begin{figure*}[h]
  \centering
  \includegraphics[width=\linewidth]{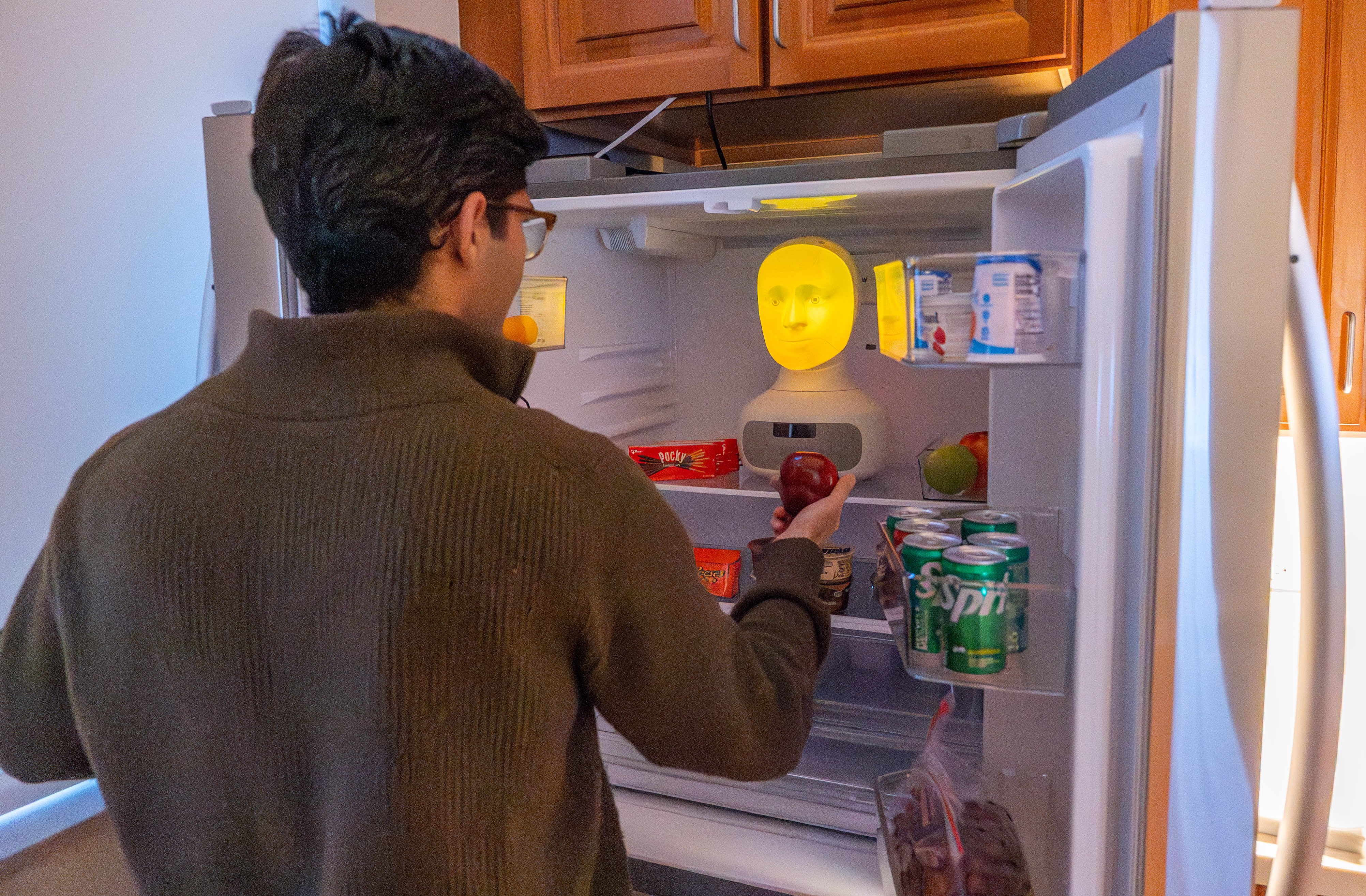}
  \caption{User Interacting with Nutrition Optimization Module (NOM).}
  \label{fig:teaser}
\end{figure*}

\section{Introduction}
Household kitchen appliances present a compelling but underexplored opportunity for supporting healthier eating behaviors. They are used frequently, embedded in daily routines, and are situated at the place and time of decision making, such as when selecting a snack or meal ingredient. Prior work on persuasive and ubiquitous computing has shown that interventions delivered at point of choice can be particularly effective, yet most domestic health technologies remain either abstract (e.g., mobile apps, speech-only conversational assistants) or episodic in use. In contrast, kitchen appliances are already deeply integrated into everyday life, making them promising candidates for unobtrusive, repeated health promotion over long periods of time. In this work, we explore how a familiar household artifact -- the refrigerator -- can be transformed into an active participant in dietary decision making through social, conversational interaction.

We frame this work within a broader vision of relational appliances: everyday domestic artifacts that function not merely as tools, but as social actors with whom people can form ongoing relationships. Unlike many anthropomorphic interfaces that leave users uncertain about how to engage, relational appliances are inherently proactive and socially legible. Their embodied presence invites interaction, their role in the household is already well understood, and their repeated use over months or years provides a natural substrate for trust, familiarity, and normalization. By embedding a physically embodied, anthropomorphic conversational agent directly into an appliance, interaction is not optional or abstract; it is encountered as part of routine activity. In the kitchen, this positions appliances as collaborators rather than passive instruments: offering suggestions, engaging in light social chat, and supporting health-related goals without requiring explicit user initiation.

Among household appliances, the refrigerator offers a particularly rich set of affordances for conversational interaction. It is located in a semi-private domestic space, and when open, its door naturally shields interaction from bystanders, reducing social awkwardness and enabling more personal exchanges. Its fixed physical layout supports the use of embedded sensors and enables spatially grounded behaviors such as gaze or gesture deixis by a humanoid robot toward food items, tightly coupling conversation with context. Refrigerator interaction is also inherently single-user, allowing the system to engage different household members individually and potentially support distinct relationships over time. While recent commercial systems such as the Samsung Bespoke AI Hybrid Refrigerator track food items and recommend recipes, their conversational assistant ("Bixby") remains a largely disembodied, task-oriented interface, passively waiting for commands and offering little social presence. In contrast, we ask: What would your refrigerator say to you if it could talk? Not as a mindless functionary, but as a social persona? We report an initial, in-home laboratory exploration of this question, examining how an embodied conversational agent embedded inside a refrigerator can influence snack choices, support health promotion, and be experienced as a social presence in everyday domestic interaction.

In our work, we target healthy snacking for behavior change \cite{HessSnack}. While some studies have found that eating many small meals is healthier than eating one large meal daily \cite{HessSnack}, there are significant associations between snacking frequency and rates of overweight and obesity among adolescents in the US \cite{snackfat}. In addition, choosing healthy snacks such as fresh fruit, and avoiding unhealthy snacks such as candy and chips, is recommended by the public health authorities in many countries, including the US \cite{snackworld}. 

In this paper, we present NOM (Nutrition Optimization Module), an anthropomorphic robotic head embedded inside a household refrigerator, designed to support in-the-moment dietary decision making through social interaction Figure~\ref{fig:teaser}. We describe a formative design study that explored the look, feel, and interaction style of a robot situated inside an appliance, leading to a set of preliminary design guidelines tailored to this unique domestic context, where issues of embodiment, privacy, agency, and social appropriateness are tightly intertwined. Building on these insights, we report findings from a controlled in-home laboratory pilot study in which NOM was deployed in a simulated everyday snacking scenario. Participants interacted with NOM during repeated refrigerator visits, where the robot either offered health-oriented food recommendations or engaged in non-persuasive social conversation. We examine behavioral outcomes, interaction dynamics, and user attitudes toward the agent, using a mixed-methods approach to assess how an embodied, relational appliance is experienced and how it may shape food choices. Together, these contributions offer an initial empirical grounding for the concept of relational appliances and highlight design considerations for embedding social agents into everyday household artifacts for home-based health promotion.

\section{Related Work}
Our work builds on and connects several strands of prior research, including smart and intelligent refrigerators, domestic human–robot interaction, persuasive health technologies, and critical perspectives on household artifacts as social and intimate technologies. While prior systems have explored informational, functional, and speculative roles for intelligent appliances, relatively little work has examined household appliances as relational social actors that engage users through embodied, situated, and time-bounded interaction at the moment of everyday decision making.

\subsection{Smart Refrigerators and Intelligent Kitchen Appliances}
Several researchers have explored the refrigerator as a site for computation, largely emphasizing information display, inventory management, and meal planning. Early research framed the “smart fridge” as a hub for tracking food contents and supporting household logistics. Luo et al. surveyed the state of the art in commercial digitally-enabled fridges and proposed extensions that could monitor food intake, suggest healthier options, and integrate with external systems such as televisions or personal computers \cite{luo2009smart}. However, this work was largely speculative and function-oriented, reflecting the technological constraints and design paradigms of the time.

Subsequent HCI studies explored user interaction with intelligent fridge concepts through graphical user interfaces and simulations. Bucci, et al. investigated interface designs for recipe recommendation and food management, focusing on usability and household coordination rather than social interaction \cite{bucci2010fridge}. Similarly, Rothensee examined reactions to simulated smart-fridge functions such as meal planning and shopping support \cite{rothensee2008user}. Notably, users in this study reacted negatively to the system’s perceived dullness, suggesting that purely functional, non-social interactions may fail to engage users. This work establishes the refrigerator as an important domestic computing platform, but treats it primarily as an informational surface or logistical tool rather than a social presence.

Early HCI work by Swan and Taylor highlighted how refrigerators function as shared domestic artifacts: sites for notes, reminders, and family coordination \cite{swan2005notes}. These studies emphasized the refrigerator’s role in everyday life and household rhythms, but did not explore interactive agents or conversational interaction. Our work extends this tradition by shifting attention from the refrigerator door as a passive display surface to the refrigerator interior as an active, conversational interaction space.

\subsection{Domestic Robots and Anthropomorphic Kitchen Artifacts}

Recent work in human–robot interaction (HRI) has begun to explore anthropomorphic robots embedded in kitchen appliances. Notably, ToasterBot examined a robotic toaster that engaged users socially while performing a familiar household function \cite{ye2023future}\cite{ye2023toaster}. This work demonstrated that anthropomorphizing mundane appliances can provoke reflection, humor, and engagement, challenging traditional boundaries between tools and social actors. However, ToasterBot focused primarily on interaction experience and meaning-making rather than health behavior change, did not examine  time-bounded interaction at the point of decision making, and was not anthropomorphically embodied.

More generally, prior HRI research has shown that embodied conversational agents can leverage social cues such as gaze, facial expression, and timing to influence user behavior and perceptions. However, much of this work situates robots in public, clinical, or explicitly interactive contexts, where users expect extended engagement. In contrast, household appliances impose severe temporal and contextual constraints: interactions are brief, opportunistic, and embedded within habitual routines. Our work contributes to domestic HRI by examining how anthropomorphism and embodiment function under these constraints, and by demonstrating that even extremely short interactions can support persuasion and relationship formation.

\subsection{Persuasive Technologies for Dietary Behavior Change}

A large literature in persuasive technology and personal health informatics has explored digital interventions to support healthier eating, including mobile apps, notifications, and context-aware reminders \cite{liu2022persuasive}. Smartphone-based \cite{hsu2014persuasive} and SMS-based \cite{kaptein2012adaptive} healthy snacking interventions have demonstrated that timely prompts can influence food choices. However, such systems rely on explicit user attention, self-reporting, or device checking, which may limit sustained engagement and increase cognitive burden.

By contrast, embedding persuasive interaction directly into a refrigerator situates intervention at the precise moment of food selection, without requiring users to shift attention to a separate device. Our findings suggest that the social presence of an embodied agent further differentiates this approach from prior mobile or GUI-based systems, enabling forms of accountability, validation, and companionship that are difficult to achieve through screens alone. At the same time, our results echo longstanding concerns in persuasive technology regarding autonomy, surveillance, and emotional impact, underscoring the need for careful design of tone, timing, and persistence.

\subsection{The Refrigerator as an Intimate and Relational Artifact}

Finally, our work resonates with more critical and reflective perspectives on domestic technologies. Blackman’s \textit{Focus on the Fridge} argues that the refrigerator’s contents constitute an intimate “archaeology” of household life, revealing habits, aspirations, and failures in ways that contrast sharply with the refrigerator’s public-facing exterior \cite{blackman2005focus}. This framing highlights the refrigerator as a uniquely personal and revealing artifact, one that mediates identity, care, and self-regulation.

By placing a social agent inside the refrigerator, NOM makes this intimacy explicit and interactive. The agent does not merely observe or display information about food; it engages users socially at a moment when private habits intersect with health ideals. This perspective distinguishes our work from prior smart-fridge systems and aligns it with emerging interest in technologies that participate in, rather than merely reflect, everyday domestic life.

\subsection{Summary}

Taken together, prior work has established the refrigerator as a promising site for computational support and has explored anthropomorphic appliances and persuasive health technologies in isolation. However, existing systems largely lack social presence, relational continuity, or situated conversational interaction at the moment of food choice. Our work bridges these gaps by introducing the concept of relational appliances and empirically examining an embodied, conversational robot embedded within a refrigerator for home-based health promotion. In doing so, we extend research on smart appliances, domestic HRI, and persuasive technology toward a more socially grounded and relational vision of everyday household technologies.

\section{Design of NOM}
The design of NOM was guided by the goal of creating a socially engaging, persuasive intervention that could operate effectively within the severe temporal, spatial, and normative constraints of a household appliance. Rather than adapting interaction paradigms from general-purpose conversational agents or mobile health applications, we treated the refrigerator as a distinct interaction context, one characterized by brief, opportunistic encounters and strong user habits. 

\subsection{Physical Design}
NOM was implemented using a Furhat Robotics Gen 2 platform\footnote{http://furhat.io}, a human-sized robotic head with a back-projected animated face and three degrees of freedom (neck pitch, yaw, and roll), selected for its ability to deliver expressive facial animation, precise gaze control, and rapid nonverbal feedback within a compact, self-contained form factor. These capabilities were particularly well suited to refrigerator-based interaction, enabling gaze deixis toward food items, subtle head nods, and expressive facial displays without requiring a full-bodied robot or additional sensors in the environment. The Furhat robot is also human-sized and positioned at user eye-level in the fridge (Figure~\ref{fig:teaser}), providing immediate engagement and an unambiguous cue that users should engage it using human-human verbal and nonverbal conversational behavior. 

\begin{figure}[h]
  \centering
  \includegraphics[width=\linewidth]{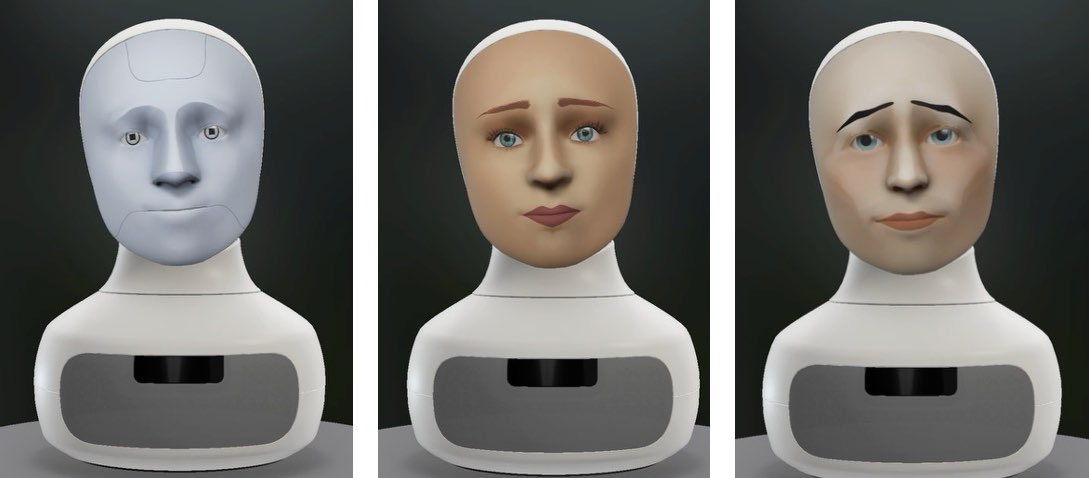}
  \caption{Some of the candidate robots who auditioned for NOM.}
  \label{fig:auditions}
\end{figure}

In order to design a robot that would be as acceptable as possible, we conducted pre-testing with 26 combinations of Furhat faces (samples shown in Figure~\ref{fig:auditions}) and available synthetic voices. The result of the online testing (136 raters, gender-balanced, recruited from Prolific\footnote{http://prolific.co}) indicated that the robot head (leftmost in Figure~\ref{fig:auditions}) was most acceptable, along with the "Kai" voice from Microsoft Azure. 

\subsection{Dialogue Design for Appliance-based Relational Interventions}
Designing a social robot embedded inside a refrigerator required rethinking conventional conversational agent interaction models. While NOM draws on established persuasive messaging strategies, including authority, scarcity, social proof, and reciprocity, early formative testing revealed a critical constraint: people are strongly habituated to closing the refrigerator door as quickly as possible. Prior literature suggests that refrigerator door-open times are typically brief and positively skewed, with averages around seven seconds \cite{elias-opentime}, and our pilot observations confirmed that users would often close the door before NOM completed even a single extended utterance. As a result, all interactions were designed to be extremely time-bounded, prioritizing immediacy and clarity over politeness and conversational depth during the core snack decision moment.

For the snack-choice scenario, we adopted a tightly structured interaction pattern. Each encounter begins with a brief, engaging greeting, immediately followed by a direct persuasive recommendation that avoids hedging, excessive politeness, or explanatory elaboration. The goal is to deliver the primary persuasive content as early as possible, before the user makes a choice or disengages. Once the user selects a snack, NOM gives a brief non-judgmental acknowledgment of whether the user followed its suggestion, to demonstrate awareness of the user's action. Only after this does NOM issue a bid for social chat, explicitly inviting conversation for those who desire it, without imposing additional interaction costs on users who prefer to disengage. Interactions conclude with a short, friendly farewell to clearly signal completion. Across all phases, NOM maintains an engaging and lightly humorous persona with a calming voice. Nonverbal behaviors, constrained by the Furhat robot’s affordances, were used strategically rather than continuously, including gaze deixis toward food items, head nods for affirmation or validation, and expressive facial displays to reinforce social presence without increasing interaction time.  

\subsection{Persuasion Principles for Appliance-based Interventions}
 NOM’s persuasive behavior was grounded in well-established social influence principles, adapted to the unique constraints of appliance-based interaction. Drawing on classic persuasive strategies, including authority, scarcity, social proof, reciprocity, and liking, we designed messages to be direct, concise, and immediately actionable, in line with the severely time-bounded nature of refrigerator interactions. Consistent with our design guidelines, persuasive content was delivered early in the interaction, without hedging or extended explanation, and framed as a confident recommendation rather than a suggestion or query. The goal was not to argue or educate, but to provide a socially grounded nudge that could influence choice within a few seconds.

Each healthy snack option was paired with a distinct persuasive framing to explore how different strategies might be received in this context. For yogurt, NOM invoked scarcity (“We’re going to run out of yogurt soon! You should grab one before they’re gone”), leveraging urgency around limited availability. For overnight oats, NOM adopted an authority frame (“As your certified nutrition robot, you should take the MUSH overnight oats. Their protein profile fits your needs right now”), positioning itself as a knowledgeable advisor. Liking and positive affirmation were used for fruit (“I love how you take care of yourself. You should grab some fresh fruit. It would be perfect for you right now”), while reciprocity framed the water recommendation (“Take some sparkling water. I pre-chilled it just for you”). Finally, social proof was employed as an alternative fruit message (“The other volunteers I’ve worked with all chose fruit. You should take some too”). Across all conditions, persuasive messages were followed by an opportunity for optional social chat and concluded with a brief farewell, preserving user autonomy while maintaining NOM’s role as a socially present, proactive collaborator rather than a passive informational system.

\section{Pilot Acceptance Study}
\begin{figure*}[h]
  \centering
  \includegraphics[width=\linewidth]{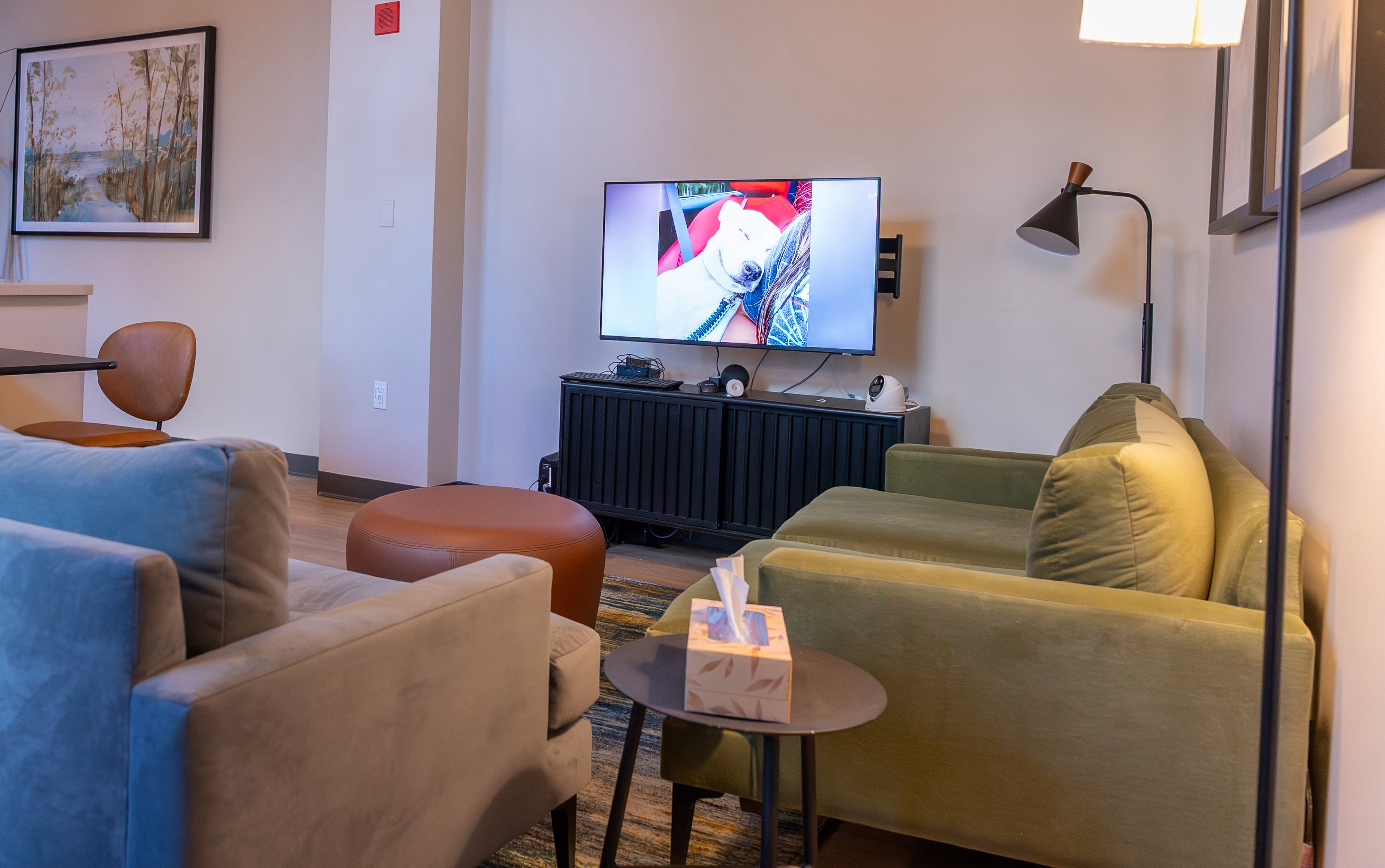}
  \caption{The Home Lab living room.}
  \label{fig:homelab}
\end{figure*}

We conducted a pilot acceptance study to evaluate initial reactions to NOM and to examine how an appliance-based relational agent is experienced during repeated, everyday snack-selection encounters. The study was carried out in the Northeastern University Home Lab, a fully furnished, one-bedroom apartment designed to closely approximate a realistic residential environment while supporting controlled behavioral data collection (Figure~\ref{fig:homelab}). The Home Lab includes a living room and adjacent kitchen outfitted with typical consumer appliances and hidden sensing infrastructure, allowing participants to interact with the refrigerator in a setting that feels domestic rather than laboratory-like. This context was intentionally chosen to balance ecological validity with experimental control.

The experiment used a between-subjects, repeated-trials design to force multiple snack-choice opportunities within a realistic and familiar activity pattern, particularly suited to an adolescent student population. Participants were seated in the living room and watched a sequence of YouTube comedy shorts on a flat screen TV display, mimicking casual at-home media consumption. Every 2–3 minutes, the video was interrupted by a simulated commercial break lasting two minutes, during which the TV displayed a “Get a snack!” prompt accompanied by a countdown timer. Participants were instructed to retrieve exactly one snack from the refrigerator during each break.

\subsection{Stimuli: NOM Implementation}
\begin{figure*}[h]
  \centering
  \includegraphics[width=\linewidth]{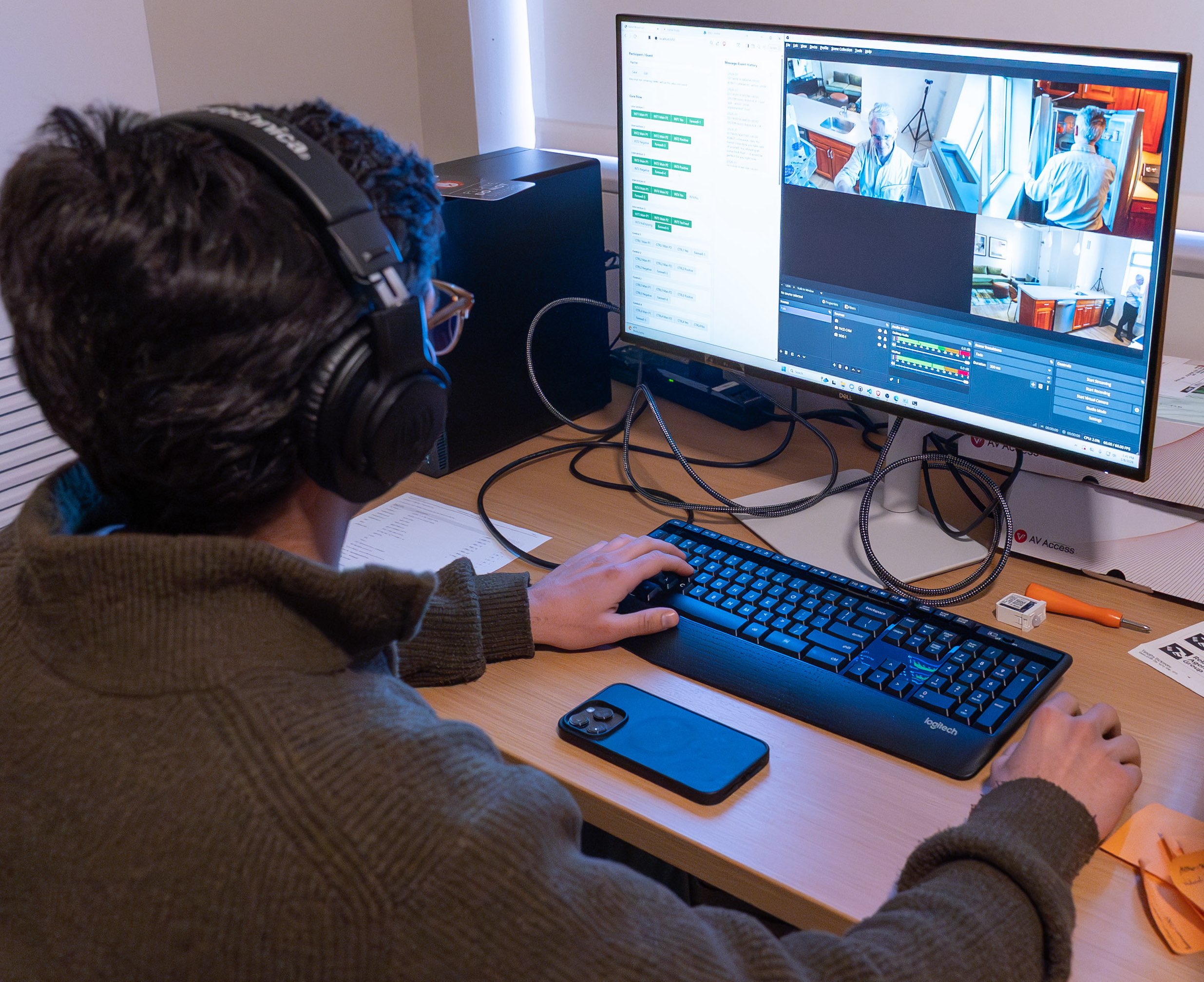}
  \caption{Wizard Of Oz control workstation.}
  \label{fig:woz}
\end{figure*}

NOM was implemented using the Furhat Gen 2 platform, positioned in the middle of a LG Model LF21G6200S 21 cu. ft. French-door refrigerator with stainless steel finish (Figure~\ref{fig:teaser}) \footnote{The refrigerator was configured to "Sabbath" mode to turn off the internal light so the robot's projected face would be visible, and disabling the open door alarm.}. NOM’s behavior during the pilot study was implemented using a Wizard-of-Oz (WoZ) control paradigm to ensure reliable timing, responsiveness, and conversational coherence within the highly time-constrained refrigerator interaction. A trained research assistant (wizard) observed and listened to participants via three video cameras positioned to capture activity in the living room and kitchen, including refrigerator interactions (Figure~\ref{fig:woz}). All NOM utterances were manually triggered by the wizard in real time. In addition to controlling the core scripted interaction sequence comprised of brief greeting, persuasive recommendation (or control message), bid for optional social chat, social chat follow-up, and farewell, the wizard had access to additional controls that allowed NOM to request clarification if a participant spoke unclearly, and to provide feedback on the participant’s snack choice. This feedback explicitly signaled whether the participant had complied with NOM’s recommendation, while maintaining a neutral and supportive tone designed to avoid berating or shaming. The WoZ setup allowed us to study user responses to an idealized version of NOM’s interaction behavior without confounds introduced by speech recognition or perception errors, while preserving the experience of a responsive, socially aware agent.

The refrigerator was stocked with food from the local market. Every single-serving item in the store that could be put in the refrigerator was cataloged and rated using the Nutri-Score system \cite{nutriscore}, which rates foods from A (best) to E (worst) on nutritional content. Seven items rated as A or B ("healthy"), and seven items rated D or E ("unhealthy") were selected, and the persuasive messages modified to recommend the healthy items.

\subsection{Experimental Design}
Participants were randomly assigned to either the INTERVENTION condition, in which NOM made health-oriented snack recommendations, or a CONTROL condition, in which NOM engaged only in non-persuasive social conversation (facts and stories about refrigerators). 

\begin{table*}[t]
\caption{Attitudes towards NOM (both conditions combined), and results of Wilcoxon signed-rank test demonstrating significant deviation from neutral scores.}
\centering
\footnotesize
\begin{tabular}{l l l c c c}
\toprule
Survey item & Anchor 1 & Anchor 7 & Mean & SD & $p$ \\
\midrule
How satisfied are you with the robot? & Not at all & Very much & 5.76 & 1.09 & $<.001$ \\
How much would you like to continue working with the robot? & Not at all & Very much & 6.24 & 0.94 & $<.001$ \\
How much do you like the robot? & Not at all & Very much & 6.19 & 0.93 & $<.001$ \\
How easy was it to interact with the robot? & Very difficult & Very easy & 5.62 & 1.20 & $<.001$ \\
How much do you trust the robot? & Not at all & Very much & 5.52 & 1.25 & $<.001$ \\
How knowledgeable was the robot? & Not at all & Very knowledgeable & 5.95 & 1.28 & $<.001$ \\
How natural was your interaction with the robot? & Not at all & Very natural & 5.00 & 1.55 & $<.05$ \\
How would you characterize your relationship with the robot? & Complete stranger & Close friend & 4.86 & 1.24 & $<.05$ \\
How annoying was the robot? & Not at all & Very much & 2.10 & 0.89 & $<.001$ \\
How persistent was the robot? & Not at all & Very much & 4.29 & 1.85 & $n.s.$ \\
\midrule
Godspeed: Anthropomorphism \cite{godspeed} (range 1-7) &  &  & 4.68 & 1.35 & $<.05$ \\
Godspeed: Animacy &  &  & 4.92 & 1.22 & $<.001$ \\
Godspeed: Likeability &  &  & 5.85 & 0.87 & $<.001$ \\
Godspeed: Perceived Intelligence &  &  & 5.74 & 1.08 & $<.05$ \\
Godspeed: Perceived Safety &  &  & 4.84 & 1.16 & $<.05$ \\
\midrule
CCR Rapport \cite{rapport} (range 1-5) &  &  & 3.82 & 0.64 & $<.001$ \\
\bottomrule
\end{tabular}
\label{tab:attitudes}
\end{table*}

\subsubsection{Measures}
In addition to timing, compliance, and other behavioral measures obtained from video recordings, participants completed post-test self-report measures including attitudes towards the robot (single items and Godspeed attitudes \cite{godspeed}, Table~\ref{tab:attitudes}), rapport with the robot (using \cite{rapport}), and measures of technology acceptance derived from the Almere/UTAUT model (\cite{utaut}). We also used the Multiple Food Test-Choice questionnaire (MFT) \cite{MFT} to evaluate the healthy food choice performance of participants through an 18-item, multiple-choice scale. The MFT questionnaire is also administered post-experiment in order to assess the effects of NOM  on changes to  participant food preference tendencies. 

\subsubsection{Participants}
Participants were recruited from our institutions' social media, and were required to be 18 years or older, speak English, and have no food allergies. The study was approved by our institution's IRB, and participants were compensated for their time.
 
\subsubsection{Study Protocol}
Upon arrival, participants were consented and then filled out baseline self-report measures on a tablet, after which they were told what they would be doing in the study. They were informed that there was a "robot in the refrigerator" but were given no additional information nor preview of NOM. Participants then sat in the living room and watched a 30-minute prepared video on the television. The video was compiled from popular YouTube comedy shorts. After approximately 2-3 minutes, a simulated "commercial break" would occur, during which the message "Get a snack!" was displayed on the TV with a two-minute countdown timer. Participants were instructed to go to the kitchen refrigerator and retrieve one snack, which they were free to consume. These breaks occurred five times during the 30-minute interval, providing five consecutive trials for each participant. Following the 30-minute video, participants completed post-session self-report questionnaires on a tablet computer and  were interviewed about their experience. Finally, they were debriefed regarding the WoZ deception before being dismissed. 

\begin{table*}[t]
\caption{Technology acceptance (UTAUT) factors for NOM. Single sample t-test comparisons against neutral value for both groups combined. Independent samples t-test between groups.}
\centering
\label{tab:TAM}
\begin{tabular}{lccccccccccc}
\toprule
& \multicolumn{3}{c}{Overall} 
& \multicolumn{2}{c}{CONTROL} 
& \multicolumn{2}{c}{INTERVENTION} 
& \multicolumn{3}{c}{Between-Group Test} \\
\cmidrule(lr){2-4}
\cmidrule(lr){5-6}
\cmidrule(lr){7-8}
\cmidrule(lr){9-11}
Item 
& M & SD & $p$ 
& M & SD 
& M & SD 
& $t$ & $df$ & $p$ \\
\midrule
Perceived Usefulness &  5.08 & 1.46 & $<.01$ & 4.03 & 1.63 & 5.73 & 0.91 & 3.08 & 19 & $<.01$ \\
Perceived Ease of Use & 5.95 & 0.99 & $<.001$ & 5.79 & 1.10 & 6.05 & 0.95 & 0.57 & 19 & n.s. \\
Perceived Enjoyment &  6.08 & 0.82 & $<.001$ & 5.74 & 1.10 & 6.29 & 0.55 & 1.55 & 19 & n.s. \\
Attitude Towards Using & 5.75 & 1.39 & $<.001$ & 5.25 & 1.49 & 6.06 & 1.29 & 1.32 & 19 & n.s. \\
Intention to Use & 5.83 & 1.25 & $<.001$ & 5.57 & 1.14 & 5.87 & 1.35 & 0.21 & 19 & n.s. \\
\bottomrule
\end{tabular}
\end{table*}

\subsection{Quantitative Results}
A total of 21 participants completed the study, 13 assigned to INTERVENTION and 8 to CONTROL. Participants were aged 20-36 (mean 25.4 SD 4.0), 71.4\% female, 90.5\% college students. Average Body Mass Index was 21.03 (SD 4.52) ("normal"). When asked whether they had "eaten healthy" over the last month participants responded 6.52 (SD 1.89) on a 1-10 scale, and when asked how important maintaining a health diet was they responded 8.43 on a 1-10 scale. 

\subsubsection{Health Behavior Compliance}
Participants in the intervention condition complied with NOM’s snack recommendations on a significantly greater proportion of trials than would be expected by chance. On average, participants followed NOM’s recommendation on 54.9\% of trials (SD = 26.4\%), compared to a 21.3\% chance baseline derived from the control condition (SD = 17.3\%). This difference was statistically significant, t(19)=3.20, p=.005, indicating that NOM’s persuasive messages influenced snack choice behavior.

Exploratory correlation analyses revealed that compliance with NOM’s recommendations was higher among participants who reported that maintaining a healthy diet was less important to them at baseline (r=-.48, p=.03). This suggests that NOM’s persuasive influence may have been strongest for participants with lower intrinsic motivation for healthy eating, a pattern consistent with prior work on just-in-time and socially mediated health interventions.

Participants in the intervention condition selected healthy snacks on a greater proportion of trials (73.1\%, SD = 18.0\%) than participants in the control condition (56.3\%, SD = 29.3\%). While this difference was in the expected direction, it did not reach statistical significance, t(19)=1.60, p=.117. A one-sample t-test comparing overall snack choices to a 50\% baseline (reflecting equal availability of healthy and unhealthy items) showed that participants selected healthy items at a rate significantly above chance, t(20)=3.22, p=.004.

Pre-post changes in the MFT measure of healthy food choice was conducted to determine whether NOM interactions led to translatable changes in food choice behavior. Participants in the CONTROL group did not change in mean MFT values from pre- to post-session (2.83 SD 0.28 to 2.83 SD 0.30), while those in the INTERVENTION group increased (2.56 SD 0.47 to 2.71 SD 0.41), although this difference was not significant. 

\subsubsection{Interaction Process}
To examine whether NOM affected decision efficiency, we analyzed refrigerator door-open times and snack selection times across all valid trials (n=59). Mean door-open duration did not differ significantly between conditions (control: 46.6 sec, SD 14.2; intervention: 43.5 sec, SD 14.2), t(88)=1.02, p=.31. However, the time to select a snack was significantly shorter in the intervention condition (mean 14.3 sec, SD 6.5) than in the control condition (mean 21.4 sec, SD 11.2), t(88)=3.78, p<.001. This suggests that NOM’s recommendations reduced deliberation time during snack selection.

Across conditions, participants engaged with NOM’s social chat bids on 89.5\% of trials (85/95). Uptake rates were high in both the control (92.5\%) and intervention (87.3\%) conditions, with no significant difference between groups, chi-squared(2)=5.5, p=.113. This indicates that participants were generally willing to engage socially with NOM, independent of persuasive content.

\subsubsection{Acceptance and Attitude} 
Across both conditions, ratings on all technology acceptance (UTAUT) dimensions were significantly above the neutral midpoint (all p<.01), indicating overall positive reception of NOM (Table~\ref{tab:TAM}). Participants in the INTERVENTION group scored NOM significantly higher on Perceived Usefulness. While all other between-group differences on individual acceptance subscales were higher in the INTERVENTION condition, the differences were not  significant.

Similarly, perceived rapport with the robot was high across both conditions and significantly above neutral, but with no significant difference between INTERVENTION and CONTROL groups. 
Measures of satisfaction, trust, perceived intelligence, naturalness, and likability were all significantly above neutral across participants. Godspeed questionnaire results indicated that NOM was perceived as moderately anthropomorphic, intelligent, and safe, with no significant differences between conditions.  These findings suggest that NOM was experienced as engaging and socially acceptable, regardless of whether it delivered persuasive health messages.

\subsection{Qualitative Results}
We transcribed 20 post-session interviews, totaling 150 minutes and 27,052 words. Transcripts were analyzed using inductive thematic analysis, focusing on participants’ experiences of interacting with NOM during snack selection, their perceptions of the robot as a social presence, and their reflections on persuasion, comfort, and desirability in a household context.

Across conditions, participants articulated rich and sometimes ambivalent reactions to encountering a conversational, anthropomorphic agent inside a refrigerator. We report five major themes that characterize how NOM was experienced and interpreted.

\subsubsection{Decision Support and Reduced Cognitive Effort}
A dominant theme was NOM’s role as a decision aid at moments of indecision. Many participants described habitual uncertainty when opening the refrigerator, often defaulting to the first visible or most convenient item. NOM’s recommendations were experienced as relieving this cognitive burden by narrowing options or making the decision for the user.  Several participants explicitly framed this as a reduction in mental effort: NOM streamlined decision making, allowing them to act quickly without deliberation. \textit{"...what I hate most is when I open my refrigerator to figure out what I'm going to eat. But with this robot, I don't have to think about it. It reduces a lot of mental effort."} (P218, INTERVENTION). This was particularly salient for participants who self-identified as indecisive or cognitively fatigued, especially during late-day snacking. Rather than perceiving this as a loss of autonomy, participants frequently welcomed the delegation of choice, describing NOM as a “food manager” or guide that helped them navigate competing options.

Importantly, NOM’s influence was not always experienced as directive. Participants emphasized that the final choice remained theirs, and that the robot’s suggestions felt supportive rather than coercive. Even when participants did not follow the recommendation, they often described the interaction as helpful for structuring their thinking or prompting reflection. This sense of assistance without obligation contributed to overall acceptance of the system.

\subsubsection{Persuasion Through Social Presence and Accountability}
Participants consistently distinguished NOM from conventional informational systems by emphasizing its social presence. Being observed, addressed, and responded to by a physically embodied agent altered how participants experienced the act of choosing food. For some, this manifested as increased awareness or self-monitoring; for others, it functioned as a form of gentle accountability.
\textit{"the face kind of really helped"} (P201 INTERVENTION).

Several participants noted that NOM’s gaze and timing made its recommendations feel personally directed, increasing their salience. In some cases, participants reported changing an intended choice—often from an unhealthy to a healthier option—specifically because the robot “noticed” them or drew attention to a particular item. This effect was amplified by the robot’s nonverbal behaviors, such as looking toward a food item or nodding in response to a choice, which participants interpreted as communicative and meaningful. This effect was even observed in the CONTROL group, when NOM did not talk about snack items at all. \textit{"wherever the robot was looking, so I was also looking at the same product. ... So, for example, it was looking at yogurt, so I thought, should I eat yogurt?"} (P104, CONTROL).

At the same time, this sense of being watched produced mixed reactions. While some participants described increased motivation to “do the right thing,” others expressed mild discomfort or concern about monitoring, particularly when they were unsure what the robot could perceive or remember. This tension highlights a core design challenge for relational appliances: social accountability can be persuasive, but it can also introduce concerns about surveillance and judgment if not carefully framed.

\subsubsection{Validation, Guilt, and Emotional Responses to Feedback}
NOM’s responses to user choices elicited a range of emotional reactions, revealing the affective complexity of food-related interaction. When NOM validated a participant’s choice, either explicitly or through nonverbal affirmation, participants frequently described feeling reassured or affirmed in their decision. This validation was particularly valued when the participant selected a food that aligned with their existing habits or health goals. \textit{"these would be the snacks that I would normally have, but it is good to know that your decisions are validated."} (P215 INTERVENTION). 

Conversely, when participants deviated from NOM’s recommendation, emotional responses varied. Some participants reported feeling mild guilt or increased self-consciousness, especially when repeatedly choosing less healthy options. Others emphasized that NOM’s nonjudgmental tone prevented negative affect, even when they did not comply. Notably, participants appreciated that the robot did not shame or scold them, and several explicitly contrasted this with how they imagined a more lecturing or moralizing system might feel. \textit{"So even though I did not pick whatever it suggested ... it didn't make me feel bad about it."} (P219 INTERVENTION). 

These findings suggest that emotional responses were shaped not only by what NOM said, but how it said it—particularly its calm voice, humorous tone, and brief utterances. The balance between prompting reflection and avoiding moral pressure emerged as an important factor in participants’ comfort with persuasion.

\subsubsection{From Initial Shock to Normalization and Relationship Formation}
Encountering a human-like face inside a refrigerator was initially surprising, and often startling, for many participants. First impressions frequently involved shock, confusion, or mild fear, with some participants describing the robot as “scary” or “skull-like” before it began speaking. However, these reactions were typically short-lived. Once NOM engaged verbally, participants reported rapid acclimation, and many described becoming comfortable within a few interactions. 
\textit{"I guess I got used to it very quickly, but it was a bit, ... surprising and shocking to see a human type face inside the fridge."} (P221 INTERVENTION).
\textit{"my impression did change. I was having genuine conversations after every time I opened the fridge, so that's a good thing, like it wasn't more machine like. ... It went better the more you use it."} (P201 INTERVENTION).

Over repeated encounters, participants described a process of normalization in which NOM shifted from an odd novelty to an expected part of the refrigerator experience. Some participants began anticipating NOM’s remarks or looking forward to opening the fridge, especially when the robot used humor or remembered prior interactions. This gradual normalization supported the development of what participants described as a “connection” or sense of relationship, even within the brief interaction windows imposed by the refrigerator context.
\textit{"later on the connection built up, so it was pretty good."} (P214 INTERVENTION).

Several participants commented that the robot could be a good social companion for those who live alone,
\textit{"If I'm ... alone by myself ... hanging around at home and trying to get something to eat. And it's kind of okay to have someone to talk to.} (P108 CONTROL), or that they would like to engage in social chat when deciding what to eat, \textit{"If I'm taking suggestions from it consciously, I would like to have some chit chat."} (P220 INTERVENTION).

At the same time, not all participants desired or embraced relational depth. A minority expressed that while the interaction was entertaining, they did not necessarily want an ongoing social relationship with a refrigerator. These participants were more likely to question the need for a human-like embodiment or to suggest alternative forms, such as voice-only or screen-based agents.

\subsubsection{Embodiment, Space, and Domestic Appropriateness}
Participants frequently reflected on the physical presence of the robot in relation to the practical constraints of the refrigerator. While many felt that the face and head substantially enhanced engagement, making the interaction feel more “human” and socially meaningful, others viewed the embodiment as intrusive or impractical. Concerns about lost storage space, cost, and visual incongruity were common. \textit{"I prefer smaller and it take less fridge space."} (P108, CONTROL).  These reflections often led participants to propose alternative embodiments, including smaller forms, abstract shapes, or cartoon-like characters. Notably, even participants who questioned the physical head frequently endorsed the function of NOM, including health guidance, reminders, or conversational support, suggesting that acceptance of relational appliances may hinge on separating social functionality from specific embodiments.

Despite these concerns, participants consistently emphasized the refrigerator’s suitability as an interaction site. The privacy afforded by the open door, the one-user-at-a-time interaction, and the tight coupling between conversation and visible food items were all cited as making interaction feel less awkward and more contextually appropriate than other domestic technologies. Several participants noted that they felt more comfortable interacting with NOM in this semi-private space than they would have in a more public or shared household location.

\subsubsection{Overall Acceptance}
When asked whether they would want NOM in their own refrigerator, 
76.5\% (13/17) responded ``yes'', 5.9\% (1/17) with ``no'', and 17.6\% (3/17)
with conditional or uncertain replies. 

\subsubsection{Summary}
Taken together, the qualitative findings indicate that NOM was experienced as more than a novelty or informational tool: it functioned as a situated social actor that shaped decision making, emotional responses, and perceptions of the refrigerator itself. Participants’ accounts reveal both the promise and the tensions of relational appliances, highlighting how social presence, persuasion, embodiment, and domestic norms interact in shaping acceptance and influence at the point of everyday health-related decisions.

\subsection{Discussion}
Our findings suggest that embedding an embodied conversational agent inside a refrigerator -- a familiar, frequently used household appliance -- can meaningfully shape snack-selection behavior while being broadly acceptable as a social presence. Participants in the intervention condition complied with the robot’s recommendations at rates substantially above chance and made snack selections more quickly, indicating that NOM functioned both as a persuasive influence and as a mechanism for reducing decision friction. Importantly, these effects emerged despite extremely brief, time-bounded interactions, underscoring the value of delivering concise, direct persuasive content at the moment of choice. Qualitative findings further suggest that NOM was experienced not merely as an information source, but as a social actor: participants described feeling guided, validated, and, in some cases, gently held accountable. The robot’s engaging persona, gaze behavior, and humor helped normalize its presence over repeated interactions, transforming initial surprise or discomfort into familiarity and, for many participants, trust.

At the same time, the results highlight key tensions inherent in relational appliances. While being “watched” by the robot increased reflection and sometimes compliance, it also raised concerns about monitoring and autonomy for some users. Similarly, although embodiment and social presence supported engagement and companionship, particularly for participants who valued social chat, others questioned the practicality of a human-like head occupying refrigerator space. These findings suggest that the effectiveness of relational appliances may depend on carefully balancing persuasion, sociality, and user control, and that a single embodiment or interaction style is unlikely to suit all users equally.

\subsubsection{Limitations}
This work represents an early-stage exploration and has several limitations. The study was conducted in a controlled home-lab environment rather than in participants’ actual homes, limiting ecological validity and the ability to assess long-term use, habituation, or behavior change beyond a single session. The sample size was small and demographically skewed toward students, which constrains generalizability. NOM was operated using a Wizard-of-Oz setup, thus the interaction quality and timing may differ from a fully autonomous system. Finally, the study focused on short-term snack choices from a fixed set of foods, rather than broader dietary patterns or sustained health outcomes.

\section{Conclusion}
This work advances the concept of relational appliances by demonstrating that everyday household artifacts can function as socially meaningful, behavior-shaping actors when designed to engage users at the right moment, in the right place, and with the right degree of social presence. Through the design and evaluation of NOM, an embodied conversational agent embedded inside a refrigerator, we show that even extremely brief, time-bounded interactions can support persuasion, accountability, and emerging relational effects. Participants’ responses suggest that the refrigerator interior constitutes a uniquely powerful interactional setting, and one that affords privacy without isolation, grounds conversation in visible material context, and enables repeated, individualized encounters over time. These findings extend prior work on smart appliances and conversational agents by shifting focus from information delivery and voice-based command execution to situated social interaction at the point of everyday decision making.

At a theoretical level, this paper contributes to HCI and HRI by offering several refinements to existing accounts of social AI in domestic settings. First, our findings challenge the assumption that relationality requires extended conversation or long-term dialogue histories. Instead, relational effects—such as validation, accountability, and trust—can emerge through ultra-short, embodied interactions tightly coupled to material action. Second, NOM illustrates how strong embodiment can reduce interactional ambiguity, countering critiques that anthropomorphic interfaces necessarily confuse users about how to engage. By clearly presenting the appliance as a social actor, NOM eliminated uncertainty about appropriate interaction while maintaining user autonomy. Third, the study foregrounds the refrigerator interior as an interactional envelope, materializing privacy and displacing embarrassment that has been widely documented in prior studies of voice assistants in shared spaces.

More broadly, this work reframes talking appliances from novelty artifacts or conversational conveniences into sites of situated governance. NOM did not merely provide information; it shaped decision speed, structured choice, and prompted reflection through a combination of social presence and environmental coupling. In doing so, it exemplifies a form of environmentally embedded persuasion, where accountability is not imposed by other people or abstract systems, but emerges from interaction with the built environment itself. This perspective extends theories of distributed cognition and everyday interaction with AI by emphasizing not only relational openness, but also asymmetry, material accountability, and the role of domestic infrastructure in shaping behavior.

Taken together, our findings suggest that future domestic AI systems may be most effective, and most acceptable, when they are designed not as general-purpose assistants or disembodied voices, but as relationally specific collaborators embedded within the routines, constraints, and affordances of everyday life. Relational appliances open a new design space for home-based health promotion and beyond, inviting HCI and HRI researchers to reconsider how sociality, embodiment, and persuasion can be productively woven into the ordinary objects that already structure daily living.

\subsubsection{Future Work}
Future work will examine relational appliances in more naturalistic, longitudinal deployments, allowing investigation of how relationships with an appliance evolve over weeks or months and how sustained exposure influences dietary habits. We plan to explore alternative embodiments and interaction modalities, such as voice-only, abstract, or screen-based agents, to better understand the tradeoffs between social presence, acceptance, and practicality. Additional work will also investigate personalization, including adapting persuasive strategies, tone, and persistence to individual preferences, goals, and household roles. Finally, moving beyond Wizard-Of-Oz control to fully autonomous sensing, perception, and dialogue will be critical for realizing the broader vision of relational appliances that can act as long-term, socially intelligent collaborators in everyday domestic life.

\section*{Acknowledgements}
Thanks for Daniela Khazen for her assistance running the study.

\bibliographystyle{ACM-Reference-Format}
\bibliography{sample-base}

\end{document}